\documentclass[pra,showpacs,showkeys, sort&compress, floatfix, square, aps, preprint]{revtex4}

\usepackage[dvips]{graphicx}
\usepackage{times}
\usepackage{textcomp}

\begin{document}

\title{GIANT MAGNETO-IMPEDANCE AND ITS APPLICATIONS}

\author{C. Tannous and J. Gieraltowski}
\affiliation{Laboratoire de Magn\'{e}tisme de Bretagne,  CNRS UMR-6135, 
Universit\'{e} de Bretagne Occidentale,  BP: 809 Brest CEDEX,  29285 FRANCE,  \\
Tel.: (33) 2.98.01.62.28,  FAX:(33) 2.98.01.73.95  E-mail: tannous@univ-brest.fr}

%\runningauthor{C. Tannous and J. Gieraltowski}
%\runningtitle{GIANT MAGNETO-IMPEDANCE AND ITS APPLICATIONS} 

\begin{abstract}
The status of Giant Magneto-Impedance effect is reviewed in wires,  ribbons
and multilayered soft ferromagnetic thin films. After establishing the 
theoretical framework
for the description of the effect,  and the constraints any material should
have in order to show the effect,  experimental work in wires,  ribbons and
multilayered thin films is described. Existing and potential applications 
of the  effect in electronics and sensing  are highlighted.
\end{abstract}



\maketitle

%\end{opening}

\section{Introduction}

Magnetoimpedance (MI) consists of a change of total impedance
of a magnetic conductor (usually ferromagnetic) under application of a static magnetic field $H_{dc}$.
When an ac current,  $I=I_{0}e^{j\omega t}$  of magnitude  $I_{0}$ 
and angular frequency $\omega $ (= $2\pi f$,  with $f$ the ordinary frequency) 
flows through the material,  it generates,  by Amp\`{e}re's Law,  a transverse 
magnetic field inducing some
magnetization. At low frequency,  the change in the transverse magnetization generates an additional
inductive voltage $V_{L}$ across the conductor: $V = RI +V_{L}$ where $R$ is the resistance.
Hence MI can be written as $Z=R+j\omega \phi/I$,  where the imaginary part is given by the ratio of magnetic flux to ac current
 and MI field dependence is related to the transverse permeability.
When frequency increases the current gets distributed near the surface of the conductor, 
changing both resistive and inductive components of the total voltage $V$. The field dependence of 
MI is dictated by skin depth $\delta_{s}=\frac{c}{\sqrt{2 \pi  \omega \sigma \mu}}$ where $c$ is 
the velocity of light,  $\sigma$ the 
conductivity and $\mu$ the permeability (For non magnetic ordinary metals,  $\mu=1$).  
The current distribution is governed not only by the shape of the conductor and frequency but also the transverse 
magnetization depending on $H_{dc}$. \\

Typically MI increases with frequency,  attains a maximum at frequencies for which
the skin effect is strong ($\delta_{s} << a$ where $a$ is a characteristic length scale such as the wire radius or ribbon/film thickness)
and then decreases since permeability becomes insensitive to the field at high enough frequency. \\

\begin{figure}[!h]
\begin{center}
\scalebox{0.3}{\includegraphics[angle=-90]{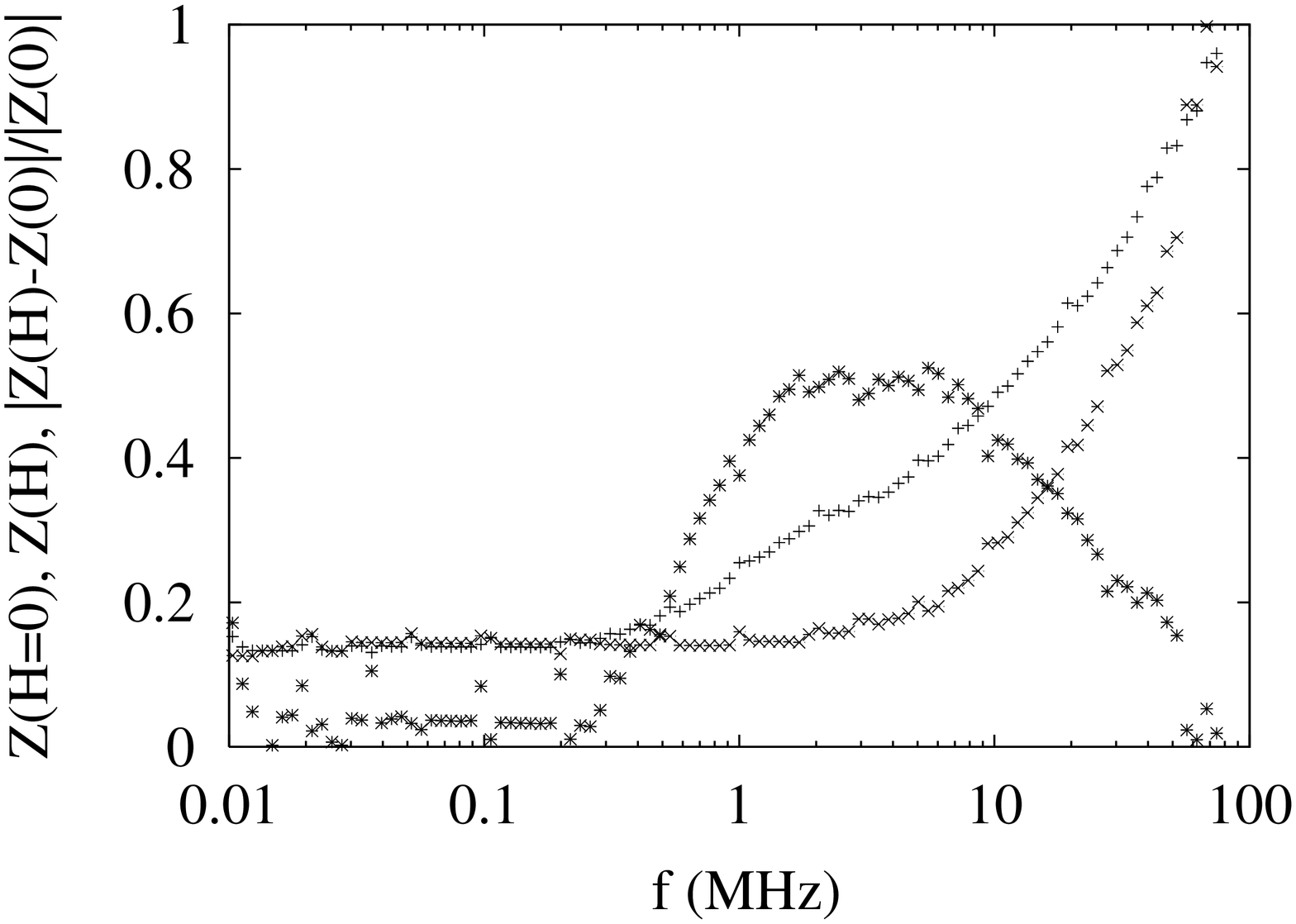}}
\end{center}
  \caption{Normalised impedance Z(H=0) and Z(H) versus frequency for amorphous Fe$_{4.3}$ Co$_{68.2}$ Si$_{12.5}$ B$_{15}$ wire
(30 micron diameter) with $H$=10 Oe. The upper monotonous curve (+) is $Z(H=0)$,  the lower monotonous curve (x) is
 $Z(H)$ and the bumpy curve (*) is the ratio $|Z(H)-Z(H=0)|/|Z(0)|$. The largest ratio reaches 60 \%. Note the finite frequency
range where the large variation is observed. The figure 
is adapted from Panina et al. \cite{panina}.}  
\label{fig1}
\end{figure}

MI effect is ordinarily weak and did not attract much attention in the past.
Interest in MI was triggered in the early 90's when Panina et al. \cite{panina} 
and Beach et al. ~\cite{beach94a}
reported a very large (Giant) MI effect in amorphous ferromagnetic FeCoSiB
wires with small magnetic fields and at relatively low frequencies (see Fig.~\ref{fig1}). This shows that  
a large variation of the MI is 
observable in a finite frequency range for reasons that will be explained in later sections.
Later on,   Machado et al. \cite{machado} observed a smaller effect in 
Fe$_{4.6}$Co$_{70.4}$Si$_{15}$B$_{15}$ thin films  and  Beach et al. ~\cite{beach94b} in ribbons. \\

The dramatic variation
of the MI (that can reach in some cases values larger than 800\%) with small magnetic fields 
(a few Oersteds) and at relatively low frequencies (tens of MHz) in 
widely available materials is the origin of the interest in the Giant MagnetoImpedance (GMI) effect. \\
 
GMI materials share the property of being magnetically soft (easy to magnetize) and are now available
as wires (typical diameter of the order of a mm),  microwires (typical diameter of the order of a micron),  
ribbons,  magnetically coated metallic (usually nonmagnetic) wires (tubes),  thin films and 
multilayers,  albeit,  the effect itself 
occurs with widely differing magnitude depending on the geometry,  the constituent materials and their layering.\\

In order to clearly identify  GMI,  several observations should be made:

\begin{enumerate}
\item A very large change (of the order of at least 100\% variation)
of the impedance should occur with an external dc magnetic field $H_{dc}$. The change expressed in \% is 
defined by the largest value of the ratio:
\begin{equation}
r(H_{dc})=100 \times |Z(H_{dc})-Z(H_{sat})|/|Z(H_{sat})|
\end{equation}
where $Z(H_{dc})$ is the impedance
measured  in the presence of the dc magnetic field $H_{dc}$ and $Z(H_{sat})$ is the impedance measured at
the saturation limit when the magnetization does not change any longer with the applied field.
\item The external dc magnetic field $H_{dc}$ should be on the order of a few Oersteds only (see Table I 
on magnetic units and quantities).
\item  The frequency range is on the order of MHz or tens of MHz (excluding any effect based
on Ferromagnetic Resonance (FMR) where the frequencies are typically in the GHz range \cite{britel}). In many materials
this means that the skin depth $\delta_{s}$ (typically microns at these frequencies) is larger than the thickness of
 the material (typically a fraction of a micron). When the frequency is in the GHz,  $\delta_{s}$ is generally very small
with respect to the thickness.  It should be stressed that in ordinary metals,  the skin depth does not
depend on permeability,  whereas in magnetic materials,  the behavior of the permeability on geometry
(see for instance section IV), temperature,  stress,  composition and so on,  is reflected in the skin depth. 
In addition, permeability might be changed by post-processing the material after growth with annealing under
presence or absence of magnetic field or mechanical stress...   
\end{enumerate}

\begin{figure}[!h]
\begin{center}
\scalebox{0.5}{\includegraphics*{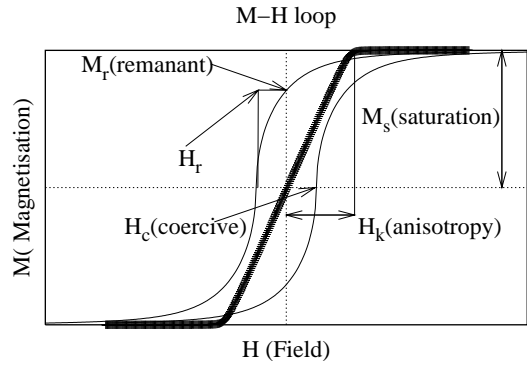}}
\end{center}
  \caption{Single domain hysteresis loop obtained for an arbitrary angle,  $\alpha$,  between the magnetic
 field and the anisotropy axis. Associated quantities such as coercive
field $H_{c}$,  remanent magnetization $M_{r}$ and field $H_{r}$ (given by the intersection
of the tangent to the loop at $-H_{c}$ and the $M_{r}$ horizontal line) are shown. The thick line
is the hysteresis loop when the field is along the hard axis ($\alpha$=90 degrees,  in this case) and
$H_{k}$ is the field value at the slope break. Quantities  such as $H_{c}$, 
$M_{r}$ and $H_{r}$ depend on $\alpha$.}
\label{fig2}
\end{figure}

GMI is a {\itshape classical phenomenon} that can be explained thoroughly on the basis
of usual electromagnetic concepts \cite{squire, knobel, valenzuela} in sharp contrast with Giant Magnetoresistance 
(GMR) effect where resistance is changed by  a magnetic field. 
GMR requires Quantum Mechanical concepts based on the spin 
of the carriers and their interactions with the magnetization of the magnetic material. 
Several general conditions must be satisfied by any material in order to show a GMI.
\begin {enumerate}
\item The material should be magnetically soft. That is,  it should be easily 
magnetised or in other words must have a relatively narrow hysteresis curve
implying,  in general,  small losses in the course of the magnetization cycle.
\item The material should have a well defined anisotropy axis. That means 
there must be a direction along which the magnetization of the material
lies on the average (easy axis). However the value of the anisotropy field $H_{k}$ 
 (see Fig.~\ref{fig2}) should be relatively small (on the order of a few Oersteds). 
The typical ratio of $H_{k}$ to $H_{c}$ must be about 20. That insures observation
 of large magnetoimpedance effects,  typically.
\item The coercive field $H_{c}$ must be small (fraction of an Oersted) and
the hysteresis loop thin and narrow. Since $H_{c}$ and the shape of the hysteresis
loop change (see Fig.~\ref{fig2}) with the angle the magnetic field makes with the easy axis (or Anisotropy
axis) of the material,  these are taken at the reference point when the field is
along the easy axis.  
\item The ac current $I=I_{0}e^{j\omega t}$,  injected in the material,  should be perpendicular to 
the easy axis (or anisotropy direction) and the magnetic field it creates $H_{ac}$ should be small with respect to 
$H_{k}$ (See Table I on magnetic units and quantities).
\item The material must have a small resistivity ($\le 100 \mu \Omega$.cm ) since it carries the ac current.
 This is important,  since many magnetic materials have large resistivities. Amorphous
metals are interesting in that respect since, typically,  their resistivities at room temperature
are in the 100 $\mu \Omega$.cm range.
\item The material should have a large saturation magnetization $M_{s}$ in order to boost
the interaction with the external magnetic field.
\item Above arguments are equivalent to a very large permeability at zero frequency 
(the ratio $M_{s}/H_{k}$ gives a rough indication for this value). This means several 1000's (see Table II).  
\item The material should have a small magnetostriction (MS). This means,  mechanical effects caused by
application of a magnetic field should be small. Mechanical stress due to MS
alters the soft properties of the material by acting as an effective anisotropy. This alters the direction
of the anisotropy,  displacing it from the transverse case and thereby reducing the value of the MI. 
Typical case materials are displayed along with their MS coefficient in Table II.
\end{enumerate}

\begin{figure}[!h]
\begin{center}
\scalebox{0.3}{\includegraphics[angle=-90]{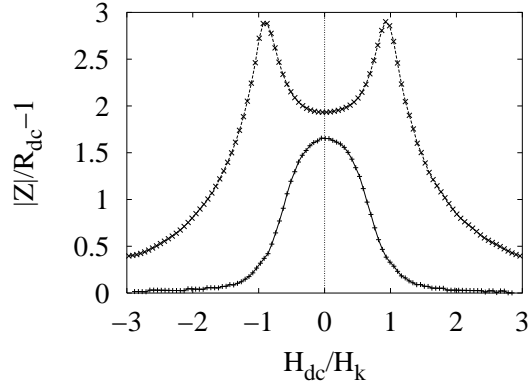}}
\end{center}
  \caption{Variation of normalised impedance $|Z|/R_{dc}$ versus magnetic field $H_{dc}$ at fixed low (single peak) and high frequency (double peak). As the frequency increases, the peak shifts toward the anisotropy field $\pm H_{k}$. That is why at low frequency a single peak is observed. The rounding of the peaks is due to a distribution of the anisotropy field direction.}
\label{fig3}
\end{figure}

The general theory of the MI effect is widely available in classic textbooks \cite{landau} (for a long cylinder)
and it has been shown experimentally that a large 
MI often occurs at frequencies of a few MHz. Changing the dc biasing field $H_{dc}$,  the maximum
 $|Z|$ can be as large as a few times the value of $R_{dc}$ the dc resistance. 
At low frequency $|Z|$ has a peak around  $H_{dc} \sim 0$ and as the frequency increases, the peak moves
 toward $H_{dc} \sim \pm H_{k}$ where $H_{k}$ is the anisotropy field. Therefore, $|Z|$ as a function of $H_{dc}$
possesses a single or a double peak as the frequency increases (Fig.~\ref{fig3}). When the direction
of the anisotropy field is well defined the peaks are sharp. \\

The behavior of $|Z|$ versus $H_{dc}$ follows very closely the behavior of the real part of the 
transverse permeability versus $H_{dc}$ as we will show in later sections on wires and ribbons. Therfore it is
very important to develop an understanding for the processes controlling the permeability. \\

Material permeability depends on sample geometry,  nature of exciting field, temperature,  
frequency,  stress distribution in the material as well as internal configuration of the 
magnetization that might be altered by processing or frequency.  
For instance,  some materials should be annealed under the presence of a magnetic field or a mechanical stress 
in order to favor some direction for the magnetization or to release the stress contained in them. \\

Regarding frequency,  when it is large enough ( $>$ 1MHz is sufficient in many materials) Domain Wall Displacements (DWD)
are considerably reduced by eddy-currents and therefore magnetization varies by rotation or switching as 
if in a single domain (see Fig.~\ref{fig2}).
As a consequence,  the rotational motion of the magnetization controls the behaviour of the permeability,  
through the skin depth. \\ 

Considering  $a$ as a typical thickness (in the case of films/ribbons) or radius (in the case of wires or microwires), 
frequencies in the tens of MHz,  lead to $\delta_{s} > a$ for the observation of GMI. 
This condition depends strongly on geometry. For instance,  $\delta_{s} > a$ in 2D structures like films 
is satisfied at much higher frequencies (GHz) than in 1D structures like wires. 
This is simply due to the optimal circular shape of wires that contains in an optimal fashion the flux while, 
simultaneously,  carrying the ac current.\\ 

In terms of GMI performance,  multilayered films (such as F/M/F where F is a ferromagnet and M is metallic
non magnetic material) are preferred with respect to single layered films since they
allow to inject the ac electric current in the metallic layer and sense the magnetic flux in neighbouring or
sandwiching magnetic layers. These can be in direct contact with the metallic layer or separated from it
by an insulator or a semiconductor. Flux closure, that increases GMI, occurs when the width of the film (transverse with
respect to the ac current) is large or that the metallic layer is entirely buried in the magnetic structure
to trap the flux. \\

The progress of GMI is thrusted towards the increase of the largest value of the ratio $r(H_{dc})$
and the sensitivity given by the derivative of the ratio with respect to the field 
(see Table III for some illustrative values). \\

This  sensitivity is simply estimated by looking at the behaviour of the permeabilty $\mu'_t$ versus $H_{dc}$
as the frequency is changed. We ought to have a large variation of $\mu'_t$ about $H_{dc} \sim 0$. This happens, in
general, at low frequencies  that is when $\delta_{s} > a$. At high frequencies, we will show in later
sections that this sensitivity is either lost or one has to go to higher  $H_{dc}$ to observe it hampering 
the use of the effect in small magnetic field detection.\\ 

Applications of GMI range from tiny magnetic field detection and sensing 
(such as magnetic recording heads) to magnetic field shielding (to
degauss Cathode Ray Tubes (CRT) monitors). The reason is that GMI materials,  being soft,  possess 
large permeabilities that are required in  magnetic shielding. \\

This work is organised as follows: In section II,  a general discussion on soft materials
is presented; in section III,  GMI in wires is described. In section IV,  GMI in
films and ribbons is dicussed. Section V highlights potential applications of GMI effect
in Electronic ans Sensing devices while Section VI bears
conclusions and perspectives of GMI.

\section{Soft ferromagnetic materials for GMI}

Soft ferromagnetic materials (SFM) used in GMI applications play a key role in power distribution,  make possible
the conversion between electrical and mechanical energy,  underlie microwave
communication,  and provide both the transducers and the active storage
material for data storage in information systems. The fingerprint of any
magnetic material is its hysteresis loop (HL) whose characteristic shape
stems from two essential properties of magnetic materials: Non-linearity
and delay between input and output signals. Non-linearity is given by the shape of the 
loop whereas the delay is given by the width of the HL. 
The quantities associated with the HL are displayed in Fig.~\ref{fig2}. In general,  magnetic materials being anisotropic
their HL width varies with the angle the external magnetic field makes 
with some given direction. The easy axis direction is defined as the direction for which the HL width
(or coercive field $H_c$) is largest. \\

The coercive field $H_c$ represents the effort to demagnetise any magnetic material. 
For instance,  hard magnetic materials exhibit large resistance 
to demagnetization and are therefore used in materials requiring
permanent magnets whereas SFM are used in devices demanding
little effort to demagnetise or remagnetise. Hard magnets are used as permanent magnets for
many electrical applications. Some rare-earth alloys based on SmCo$_5$ and 
Sm(Co, Cu)$_{7.5}$ are used for small motors and other applications requiring an extremely
 high energy-product magnetic materials. Fe-Cr-Co alloys are used in telephone receivers and Nd-Fe-B is used for 
automotive starting motor. \\

In contrast,  SFM possess a narrower loop than hard materials and the area within the hysteresis curve
is small. This keeps the energy losses small,  during each magnetic cycle,  in 
devices based on these materials.\\

In order to qualify the different requirements for read heads and storage media in magnetic
recording,  an area belonging to the application portfolio of SFM (besides sensing and shielding), 
Table IV displays the desired characteristics. \\

SFM have high permeability (i.e. easily
magnetized) with a low coercivity (i.e. easily demagnetized). Examples 
of soft ferromagnetic materials include Fe alloys (with 3 to 4\% Si) used in motors
and power transformers and
generators. Ni alloys (with  20 to 50\% Fe) used primarily for high-sensitivity 
communications equipment. This illustrates the versatility of SFM applications that range
from mechanical to electrical and from power to communications systems. \\ 

SFM exhibit magnetic properties only when they are
subject to a magnetizing force such as the magnetic field created when
current is passed through the wire surrounding a soft magnetic core. They 
are generally associated with electrical circuits
where they are used to amplify the flux generated by the electric currents.
These materials can be used in ac as well as dc electrical circuits.

Several types of SFM exist:
\begin{enumerate}
\item Soft mono and polycrystalline ferrites
\item Powder composite magnetic materials
\item Rapidly quenched ferromagnetic materials
\item Amorphous magnetic materials
\item Nanocrystalline magnetic materials
\end{enumerate}

The soft ferromagnetic behavior in these materials arises from a spatial
averaging of the magnetic anisotropy of clusters of randomly oriented
small  particles (typically $<$ fraction of a micron). In some cases,  the 
MS of these materials is also reduced to near zero.

Metallic glasses obtained by rapid quench (e.g. splat-cooling) are  magnetically very soft.
This property is used in multilayered metallic glass power transformer cores.

Amorphous magnetic alloys such as CoP first reported in 1965 and splat-cooled materials
with attractive soft ferromagnetic properties are based either on 3d transition metals
(T) or on 4f rare-earth metals (R). In the first case,  the alloy can be stabilized 
in the amorphous state with the use of glass forming elements such as boron,  
phosphorus and silicon: Examples include Fe$_{80}$ B$_{20}$,  Fe$_{40}$ Ni$_{40}$P$_{14}$ B$_{6}$, 
and Co$_{74}$ Fe$_{5}$ B$_{18}$ Si$_{3}$ (T$_{x}$ M$_{1-x}$, with $15 < x < 30$  at \%,  approximately).
The transition metals of late order (TL where TL=Fe,  Co,  Ni) can be stabilized in 
the amorphous state by alloying with early order transition metals (TE) of 4d or 5d type (TE= Zr,  Nb,  Hf): 
some examples are Co$_{90}$ Zr$_{10}$ ,  Fe$_{84}$ Nb$_{12}$ B$_{4}$ ,  
and Co$_{82}$ Nb$_{14}$ B$_{4}$ (TE$_{1-x}$ TL$_{x}$,  where  $x$  is roughly 
 $5 < x< 15$ at \%). \\

Nanocrystalline soft magnetic materials are derived from crystallizing
amorphous ribbons of a specific composition such as the (Fe, B) based alloy family.
This class of materials is characterized by 10-25 nm sized grains of a
body-centered-cubic (Fe, X) phase consuming 70-80\% of the total volume,
homogeneously dispersed in an amorphous matrix. \\

Two families of alloys show the best performance characteristics and have emerged
as the best candidates to major SFM applications: Fe-Cu-Nb-B-Si (the "Finemet" family,  
see Table III) and
Fe-Zr-(Cu)-B-(Si) (the "Nanoperm" family). The Finemet family is
characterized by an optimum grain size of about 15 nm,  exhibits good 
properties at high frequencies and is comparable to some of the best
(and relatively costly)  Co based amorphous materials. On
the other hand,  the grain sizes consistent with optimum performance are
larger,  around about 25 nm,  in the Nanoperm family. The distinguishing
feature of the Nanoperm family of alloys is the very low energy loss
exhibited at low frequencies (60 Hz),  offering the potential for application
in electrical power distribution transformers. \\

A typical amorphous alloy with a small MS coefficient 
$\lambda_{s}$,  is Co$_{70.4}$ Fe$_{4.6}$ Si$_{15}$ B~$_{10}$. It is obtained by
alloying FeSiB that has a positive $\lambda_{s}= 25. \hspace{1mm} 10^{-6}$ with CoSiB that has
a negative $\lambda_{s}= -3. \hspace{1mm} 10^{-6}$ coefficient. Some materials that result from
this alloying can reach a very small value  of  $\lambda_{s}$. For instance, 
(Co$_{0.8}$ Fe$_{0.06}$)$_{72.5}$ Si$_{12.5}$ B $_{15}$ has $\lambda_{s} \sim -10^{-7}$ 
that can be considered as  zero since typical values of $\lambda_{s}$ are units or tens of $10^{-6}$.
This was obtained by varying systematically the concentration $x$ in the compound 
(Co$_{1-x}$ Fe$_{x}$)$_{72.5}$ Si$_{12.5}$ B$_{15}$. Starting with $x=0$ the MS
coefficient $\lambda_{s}= -3.0 \hspace{1mm} 10^{-6} $ decreases steadily to a near zero value 
$\lambda_{s}= -10^{-7} $ for $x=0.06$. Let us indicate, at this point, that a commercial
product exists called Vitrovac 6025~\textregistered with the composition Co$_{66}$~Fe$_{4}$~Mo$_{2}$~Si$_{16}$~B$_{12}$ 
that exhibits a value \cite{murillo} of $\lambda_{s}= -1.4 \hspace{1mm} 10^{-7}$. \\

A small negative value of $\lambda_{s}$ is exploited in wires, as discussed in the next section,
produces a magnetization profile  that is circular in a plane perpendicular to the wire axis whereas a positive 
$\lambda_{s}$ produces a radial magnetization profile.

\section{GMI in wires}

The search for GMI in soft magnetic wires and microwires is a topic of interest related to
possible applications as tiny field sensors. \\

At low frequencies (weak skin effect or large skin depth),  the first order expansion term  of the impedance
 $Z=R(\omega)+jX(\omega)$
versus frequency is responsible for the voltage field dependence. This term is represented 
by an inductance,  which is
proportional to the transverse permeability. When the skin effect is strong,  the total 
impedance including resistance $R$ and reactance $X$ is field dependent through the penetration depth. \\

The general theory of the MI effect in a long cylinder is widely available 
in classic textbooks \cite{landau} and it has been shown experimentally that a large 
MI often occurs at frequencies of a few MHz. Changing the dc biasing field $H_{dc}$,  the maximum
$|Z|$ can be as large as a few times the value of $R_{dc}$ for amorphous wires. 
Following Beach et al. \cite{beach94b} the behaviour of the permeability can be understood on the basis of a
simple phenomelogical model based on a single relaxation time $\tau$ for the magnetization that yields for the permeability:
\begin{equation}
\mu_t(\omega,H_{dc})=\mu'_t + j \mu"_t ; \hspace{5mm} \mu'_t=1+\frac{4 \pi \chi_0(H_{dc})}{1+\omega^{2} \tau^{2}}, \hspace{2mm}
\mu"_t=\frac{4 \pi \chi_0(H_{dc}) \omega \tau}{1+\omega^{2} \tau^{2}}
\end{equation}

The behaviour of $\mu'_t$ depicted in Fig.~\ref{fig4} indicates that the strongest variation of  $\mu'_t$ with
$H_{dc}$ happens at low frequencies as required for a sensitive sensor whereas the variation gets totally
smeared out at large frequencies. It also illustates the existence of a finite frequency range for the GMI effect
as already observed in Fig.~\ref{fig1}. The range is determined by the inverse time 1/$\tau$ that controls
relaxation processes of the transverse magnetization.\\ 

While GMR is usually  attributed to the differential scattering of conducting electrons
 whose spins make particular angles with
the local magnetization of different scattering centers,  it was believed,  like GMR,  this 
effect also resulted from the electron scattering by ac current-induced domain wall oscillations. 
Many different models exist for the description of permeability of metallic samples. Such models  
start with a uniform dc permeability throughout the sample, and then consider a class of domain 
structures with a magnetization process of DWD or domain magnetization rotations
(DMR). In the case of ac impedance,  standard
quantitative models are used without considering domain structures. Therefore, 
some current explanations of MI have to be made based on usual considerations with some additional 
modifications to include concepts of DWD,  DMR,  ferromagnetic resonance, and magnetic relaxation.  
For a straight wire of radius $a$,  conductivity $\sigma$ and permeability $\mu$
(see Fig.~\ref{fig5}) the expression for impedance is given by \cite{landau}:
\begin{equation}
\frac{Z}{R_{dc}}=\frac{R+jX}{R_{dc}}=\frac{ka}{2}\frac{J_{0}(ka)}{J_{1}(ka)}
\end{equation}
where $J_{i}$ is the i-th order Bessel function and $k=(1+j)/\delta_{s}$ where
the skin depth is given by $\delta_{s}=\frac{c}{\sqrt{2 \pi  \omega \sigma \mu}}$.
$c$ is the speed of light and $\sigma$ the conductivity. \\

Since the ac current is applied along the wire axis perpendicularly to the anisotropy (Fig.~\ref{fig5}), 
it is again the transverse permeability $\mu$ that is considered here.
In order to show that the behavior of $Z$ is dictated by the permeability, let us consider firstly
the low frequency case, i.e. $ ka << 1$. Expanding the Bessel functions, yields:
\begin{equation}
\frac{Z}{R_{dc}}=1 + \frac{1}{48}{(\frac{a}{\delta_s})}^{4}-j\frac{1}{4}{(\frac{a}{\delta_s})}^{2}
\end{equation}

In the opposite high frequency case  ($\delta_{s} << a$ or $ ka >> 1$), taking $ a \sim $ 1mm whereas
 $\delta_{s} \sim 1 \mu$ m,  we can expand the Bessel function to obtain:
\begin{equation}
\frac{Z}{R_{dc}}=(1+j) \frac{a}{2 \delta_s} 
\end{equation}

This indicates that $\frac{Z}{R_{dc}}$ could reach a value of several 1000's at high frequencies. The same result
is obtained in the case of microwires by rescaling all lengths (since we have,  in this case,   $a \sim 1 \mu$ m).
It should be pointed out that since the ac current is along the wire axis,  the magnetic field produced
is circular along or against the rotational magnetization profile (see Fig.~\ref{fig5}). Note that at the frequencies 
of interest ( $>$ 1MHz),  where we have the largest sensitivity to $H_{dc}$, eddy-currents heavily damp DWD. \\

In wires made of materials having a small negative $\lambda_{s}$, the magnetization and anisotropy 
field $H_{k}$ run in a circular fashion in a plane transverse to the ac current as illustrated 
in Fig.~\ref{fig5}. The dc field $H_{dc}$ acts to realign the magnetization along its direction,  
decreasing therefore the transverse permeability. Therefore,
 a positive $\lambda_{s}$ cannot
be used for GMI since the magnetization profile is rather radial.

The skin depth dependent classical formula does not account for the presence of domain walls that can alter the scattering
of the carriers. Panina et al. \cite{panina} and  Chen et al. \cite{chen} accounted
for domain wall scattering and showed that the latter is efficient when the inter-domain distance $\lambda$ (see Fig.~\ref{fig5}) to the
wire radius $a$ ratio $\lambda/a \le 0.1$ for frequencies $\sim$ 1MHz or greater.  
Introducing a parameter $\theta=a \sqrt{\sigma \mu \omega}$,  
they argue that a change of MI by a factor of 10 (assuming a change of  $\theta^2$ by a factor between 7 and 700) yields a change
of $\mu/\mu_{0}$ by a factor of 100 to 10, 000. Using a resistivity $\rho=1/\sigma=$125  $\mu\Omega .cm$ 
and 2$a$=0.124 mm,  the frequency $f \sim 3.1$ MHz yields a factor of 4 to 10 change in $|Z|$.
Chen et al. \cite{chen} proved also that the classical limit is recovered when $\lambda/a \le 0.01$.
The interpretation of a single or a double peak in the MI versus $H_{dc}$ is interpreted as resulting from
scattering by DMR. The scattering increases as $H_{dc}$ gets closer to the anisotropy field $H_{k}$ and then
decreases after reaching its maximum for $H_{dc} \sim \pm  H_{k}$. Wires with very small anisotropy possess a single
peak since the scattering maximum occurs at $H_{dc} \sim 0$. Frequency effects produce the same result as seen
in Fig.~\ref{fig3}. 

In the case of GMI effect in coated tubes such as the electroplated FeNi,  FeNiCo,  and CoP microwires,  Kurlyandskaya
et al. \cite{kurly} found a huge enhancement factor approaching 800\% in Fe$_{20}$Co$_{6}$Ni$_{74}$ layers (1 $\mu$ m-thick) 
electroplated onto CuBe nonmagnetic microwire (100 $\mu$m diameter) at a frequency of 1.5 MHz. That shows the 
way for producing the largest GMI in these systems.

\section{GMI in ribbons and thin films}

Sputtered or otherwise produced films or ribbons possess several advantages with respect to wires
due to the possibilities of size reduction and increase of efficiency.  Nevertheless,  oblong or elongated 
ribbon or thin film geometries are  preferred in order to emulate the wire case where the largest GMI ratios were obtained. \\

Since the main ingredients of GMI are a metal that carries an ac current  
and a nearby magnetic material that must sense strongly the flux,  2D materials or multilayers
offer more flexibility than wires and experimentally,  several excitation configuration become 
possible when we are dealing with a ribbon or a film. \\
 
The general MI measurement configuration (longitudinal,  perpendicular,  transverse) 
in ribbons and films is shown in Fig.~\ref{fig6}.
In films and ribbons,  the domain wall configuration is similar to what goes on in wires in the
sense domain walls are transverse with respect to the ac current. In Fig.~\ref{fig7} neighbouring domains 
have magnetizations pointing in different directions
and the magneto-impedance effect result from averaging over transverse domains like in the wire case. \\

When a probe current $I_{0}e^{j\omega t}$ 
is applied to a film of thickness $2a$,  the impedance is written as:
\begin{equation} 
\frac{Z}{R_{dc}}= jka \hspace{1mm} \coth(jka), 
\end{equation}

where $R_{dc}$ is the dc resistance of the film and $k=(1+j)/\delta_{s}$ with $\delta_{s}$ the skin depth
given by, as before, $\delta_{s}=\frac{c}{\sqrt{2 \pi  \omega \sigma \mu}}$. 
Since the current is applied (in the ribbon case) along the ribbon axis,  again the transverse permeability
is considered. As in the wire case, it is possible to expand $\frac{Z}{R_{dc}}$ at low (or $ka<<1$) and high frequencies
(or $ka>>1$) and show that $\mu_t$ controls its behavior with $H_{dc}$. At low frequencies ($ka<<1$), we get:
\begin{equation} 
\frac{Z}{R_{dc}}= 1 - \frac{2j}{3}{(\frac{a}{\delta_s})}^{2} 
\end{equation}

whereas, at high frequencies ($ka>>1$):
\begin{equation} 
\frac{Z}{R_{dc}}= 1 - j \frac{a}{\delta_s} 
\end{equation}

The domain wall configuration (see Fig.~\ref{fig7}) permits to estimate the permeability in the
 case of weak anisotropy field ($H_{k}$ small but finite). It should be stressed that domains,  
$H_{k}$ and permeability are all transverse.
When $H_{dc}$ is longitudinal,  it plays the role of a hard axis field since it is acting on the magnetization,  
to rotate it,  hence decreasing the transverse permeability,  similarly like in wires. \\

In order to increase the MI response,  multilayers are preferred to single layered films/ribbons.
If a sandwiched metallic layer carries the ac current that creates the flux in the neighbouring
magnetic layers,  the MI effect is considerably enhanced (Table III). Moreover,  when the 
width $2b$ is large (Fig.~\ref{fig7}) the ac flux loop is closed and stray magnetic fields small \cite{paton}. \\

Multilayers of Permalloy (Ni$_{81}$Fe$_{19}$ denoted as Py) and Ag of the type
$[$Py(x \AA )/Ag (y \AA)$]_{n}$ with thickness x = 107 \AA \hspace{1mm} and 
88 \AA,  y= 7 \AA \hspace{1mm} and 24 \AA \hspace{1mm} and
the number of layers  n=15 and 100 were grown by Sommer et al. \cite{sommer99} with magnetron 
sputtering techniques. The case n=100 displayed a GMI effect of the order of 5 to 25 \%.
The thickness of the Py layer has a strong effect on the coercive field and consequently on 
the softness of the material.
The coercive field $H_{c}$ varies from 0.85 Oersted for x=88 \AA \hspace{1mm}to reach 13.6 Oersted for
x=5000 \AA. As the Ag thickness varied from 7 to 24 \AA \hspace{1mm} the MI effect changed 
from single peak to double peak structure in a fashion akin to the effect of frequency.
Sommer and Chien \cite{sommer95} showed also that in the ribbon case,  a double or a single
peak structure in the MI results from the direction of the applied magnetic field $H_{dc}$.
When it is longitudinal (along the ribbon axis,  see Fig.~\ref{fig7}) a double peak at 
$H_{dc} \sim \pm  H_{k}$is observed,  but when it is transverse the demagnetization factor 
perpendicular to the ribbon axis becomes important. That is responsible of an apparent 
{\itshape longitudinal} permeability that gives  birth to a single peak MI.   
The ribbon MI is similar to the wire case,  however MI in single layer films remains small and it is still an open 
problem to find ways to increase it dramatically like in coated microwires. Nevertheless,  when multilayers are used
(see Table III) the GMI ratio may attain 700\%. This enhancement is due to the electrical insulation of the metallic 
layer carrying the ac current by sandwiching it with SiO$_2$ layers,  the outer magnetic layers embodying the
ac inductance created by the current.

\section{Applications of GMI in Electronics and Sensing}

Since GMI uses soft magnetic materials possessing a large permeability,  the first
immediate application is in devices related to magnetic shielding since it requires
the soft properties of the material. A host of other applications
involves the GMI per se. \\

The first of these applications are the detection of very small magnetic fields.
In order to have a general idea of the scale of current magnetic Fields 
Table V displays some orders of magnitude of natural and artificial fields. \\

Detection of magnetic field is also important and  magnetic field sensors (see Hauser et al. 
\cite{hauser}) are 
broadly classified in three categories (see Table VI):
\begin{enumerate}
\item Medium to large field detection by Hall and magnetoresistive devices. 
\item Small  to medium magnetic field detection by magnetoimpedance and 
Flux Gate sensors.
\item Very small to small magnetic field detection by SQUID's (Superconducting QUantum Interference 
Devices).
\end{enumerate} 

The possible devices are recording read heads,  magnetic guidance devices in vehicles,  boats and planes 
(with or without GPS,  i.e. Global Positioning System),  brain imaging (magneto-encephalogram or MEG devices)
and heart mapping (magneto-cardiogram or MCG devices) etc... \\
The detection of the Earth magnetic field has a host of applications for instance
in Petroleum or minerals exploration or in shielding used for Degaussing of High performance CRT monitors... \\
The general features required for sensors are not only high sensitivity,  flexibility, 
large bandwidth and lost cost but linearity is also a desired feature. 
In order to increase the sensitivity of GMI devices with respect to the dc magnetic field, 
devices were developed \cite{kim} possessing non-symmetric variation of the MI with respect to $H_{dc}$.
The materials used are field annealed Co-based amorphous ribbons 
(Co$_{66}$ Fe$_{4}$ Ni\hspace{1mm}B$_{14}$ Si$_{15}$) (see Table III). The asymmetry
of the MI profile allows a very sensitive detection of the magnetic field specially around $H \sim 0$ and when
the asymmetry of the profile is so sharp that it becomes step-like,  we have a so-called "GMI valve" based device.
In these devices,  the sensitivity might be enhanced to reach 1000\%/Oe (compare with values in Table III).
Asymmetry may also be induced by torsional stress such as in wires of Fe$_{77.5}$Si$_{7.5}$ B~$_{15}$ \cite{ryu}.
Altering the GMI response with mechanical stress paves the way toward the development of strain sensors
that can be used in several areas of engineering and science.

\section{Conclusion}
\begin{figure}[!h]
\begin{center}
\scalebox{0.3}{\includegraphics[angle=-90]{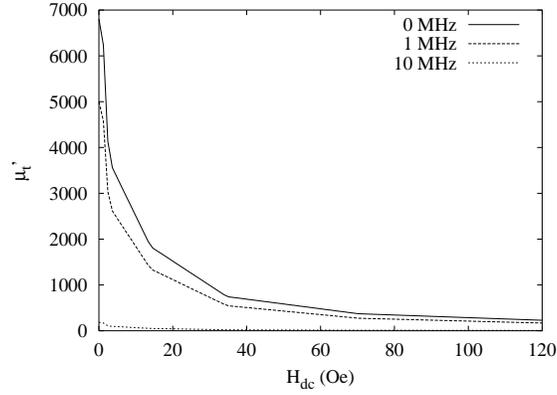}}
\end{center}
  \caption{Variation of the transverse permeability versus magnetic field $H_{dc}$. The physical quantities used are the same of 
Beach et al. \cite{beach94a} such as the experimentally determined function $\chi_0(H_{dc})$ and the inverse relaxation
 time $\frac{1}{\tau}=10.45$ MHz. The latter determines the range of variation of interest for the permeability.}
\label{fig4}
\end{figure}

\begin{figure}[!h]
\begin{center}
\scalebox{0.3}{\includegraphics[angle=-90]{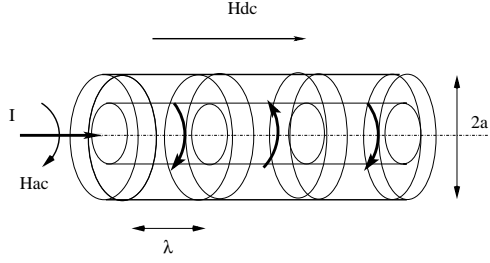}}
\end{center}
  \caption{Bamboo-like domain wall structure in a cylindrical wire with magnetization profiles
counter-rotating about a central core where the ac current $I$ circulates.
 It is the small negative value of MS $\lambda_{s}$ 
that induces a circumferential residual anisotropy through the inverse magnetostriction 
effect  \cite{panina} producing the circular magnetization profiles. 
The resulting circular anisotropy axis (easy axis),  along the magnetization is 
perpendicular to the direction of the ac current.
 Therefore,  the dc field $H_{dc}$ plays the role of a hard axis that damps magnetization and
consequently decreases the permeability. A positive $\lambda_{s}$ results in a radial magnetization profile that cannot be exploited in GMI applications.}
\label{fig5}
\end{figure}

\begin{figure}[!h]
\begin{center}
\scalebox{0.3}{\includegraphics[angle=-90]{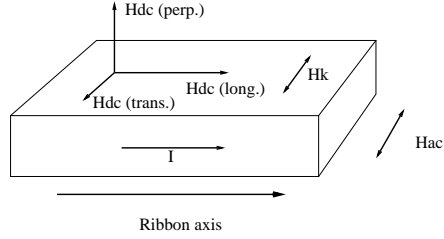}}
\end{center}
  \caption{General magnetoimpedance measurement setup in an elongated ribbon. The static field 
\textbf{Hdc} might be longitudinal,  transverse or perpendicular to the ac current 
$I=I_{0}e^{j\omega t}$ that creates the alternating transverse \textbf{Hac}. $H_{k}$ is the anisotropy field,  transverse 
to the ac current direction. When \textbf{Hdc} is longitudinal,  it plays the role of a hard axis field
like in wires that decreases the transverse permeability.}
\label{fig6}
\end{figure}

\begin{figure}[!h]
\begin{center}
\scalebox{0.3}{\includegraphics[angle=-90]{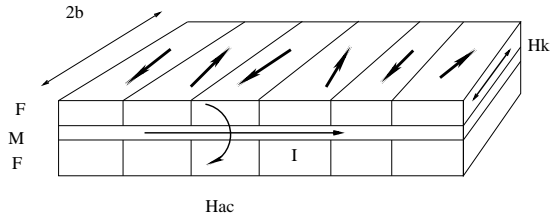}}
\end{center}
  \caption{Domain wall configuration in a multilayered structure. The magnetization direction is shown in every domain.
 The ac current flows in the sandwiched metallic layer,  producing an ac flux in the 
surrounding magnetic layers.}
\label{fig7}
\end{figure}

Good candidates for GMI are amorphous Cobalt rich ribbons,  wires,  glass covered microwires
and multilayers made of a metallic layer sandwiched between ferromagnetic materials
(with or without intermediate insulating layers). \\
Zero-field annealing or annealing under a magnetic field or with the application of some mechanical stress 
favors orienting the magnetization along a desired direction. In order to get a fast reduction
of the transverse permeability (see Fig.~\ref{fig4}) under the application of an external field $H_{dc}$, there must be an optimal
distribution of the magnetization around the desired direction \cite{panina} and that
can be obtained with suitable growth or annealing conditions. Reduction of the stress contained 
into the as-grown materials by sputtering or quenching can also be obtained with annealing.\\
The current activity in GMI studies is oriented toward the development of
devices using a built-in magnetic field rather than an applied external 
field. This is reminescent of the p-n,  Schottky,  heterojunctions etc...
where the built-in field is electrical. \\
The discovery of magnetic bias produced a revolution in magnetism because 
of the many potential applications in storage,  sensing,  spintronics... 
Nevertheless,  there are other ways to produce built-in magnetic fields. For instance Ueno et al. 
\cite{ueno} produced susch an internal field by superposing two sputtered films
of Co$_{72}$ Fe$_8$ B~$_{20}$  with crossed anisotropy axes,  i.e. with the 
anisotropy axis of the bottom layer making angles of opposite sign with some common direction.  \\
While GMI in wires and ribbons is steadily progressing,  the case of single layered films
is still lagging behind in favor of multilayered films. Ways must be developed
in order to increase the sensitivity of single or multilayered structures in order
to reach the level observed in wires and microwires.
The range of applications will substantially explode once GMI in thin films will be competitive
with wires and ribbons. \\
   
{\bf Acknowledgement} \\
The  authors wish to acknowledge correspondance with P. Ripka and
his kind providing of re(pre)prints of his work as well as friendly 
discussions with A. Fessant regarding characterisation and impedance 
measurements. \\

\newpage

\begin{center}
{\bf FIGURES AND TABLES}
\end{center}

\begin{table}[!h]
\begin{center}
\begin{tabular}{|l|c|r|}
\hline
 Physical Quantity & SI &  CGS\\
\hline 
  Bohr magneton $\mu_{B}$ &   $0.927 \hspace{1mm} 10^{-23}$A.m$^{2}$ &   $0.927 \hspace{1mm} 10^{-20}$emu \footnote{also expressed in erg/Oe}\\  
\hline
   Vacuum permeability $\mu_{0}$                 &   $4\pi  \hspace{1mm} 10^{-7}$ V.s/A.m &        1  \\
\hline
   Field Strength H &   A/m & 4$\pi  \hspace{1mm} 10^{-3}$ Oe\\
   Example &   80 A/m & $\sim$ 1Oe\\
\hline 
    Polarisation or magnetization        &   $\mu_{0}M_{s}$ &  $ 4 \pi M_{s} $    \\
    with saturation  value  $M_{s}$         &                 &                      \\
    Example  $M_{s}$\footnote{It is also expressed in Gauss with 1Gauss= 1$emu/{cm}^3$ }    & 1A/m &  $4\pi 10^{-3}$ emu/cm$^{3}$\\
\hline 
   Induction  $B$ &   $B=\mu_{0}(H+M)$ &   $B=H+4 \pi M$\\
   Example                 &   1 Tesla= 1V.s/m$^{2}$ &   10$^{4}$G\\
\hline
     Susceptibility                &   $M=\chi H$ &  $M=\chi H $    \\
     Example                &   $\chi=4\pi$ &  $\chi =1$    \\
\hline
     Energy density of magnetic Field                &   $BH/2$ &  $BH/8\pi$    \\
     Example               &   1J/m$^{3}$ &  10erg/cm$^{3}$    \\
\hline 
     Energy of magnetic matter    &   $\mu_{0}HdM$ &  $HdM$    \\
      in an external field        &                                &     \\
\hline
     Anisotropy constant K         &      10$^{5}$ J/m$^{3}$     &    10$^{6}$ erg/cm$^{3}$    \\
\hline
    Anisotropy Field     &   $H_{K}=2\frac{K}{\mu_{0}M_{s}}$ &  $H_{K}=2\frac{K}{M_{s}}$     \\
    Example         &       $10^{6}$ A/m            &     $4\pi {10}^3$ Oe    \\
\hline
    Exchange Field      &   $\frac{J}{\mu_{B}}$ &  $\frac{J}{\mu_{B}}$     \\
    Example         &       ${10}^9$ A/m          &     $4\pi {10}^6$ Oe    \\
\hline
    Demagnetising field in a thin film    &   $-M_{s}$ &  $-4\pi M_{s}$    \\
\hline 
    Energy density of     &   $\mu_{0}M_{s}^{2}/2$ &  $2\pi M_{s}^{2}$    \\
    Demagnetising field in thin film        &                                &     \\
\hline
    Magnetostriction      &                   &        \\
    coefficient         &     $\pm \lambda_{s} \sim 10^{-6}$  &  $\pm \lambda_{s} \sim 10^{-6}$   \\
\hline     
\end{tabular}
\end{center}
\caption{Correspondance between magnetic Units in the SI and CGS unit systems. Note that magnetic field units are A/m and Oe and 
Induction's are Tesla and Gauss. In vacuum or non-magnetic materials in CGS values in Oe and Gauss are same.}
\label{tab1}
\end{table}

\begin{table}[!h]
\begin{center}
\begin{tabular}{|c|c|c|c|}
\hline
 Alloy & $H_{c}$ (mOe)& $\mu_{max}$ at 50 Hz  & magnetostriction \footnote{It is defined as $\lambda_{s}=\delta l/l$,  the 
largest relative change in length due to application of a magnetic field} Coefficient \\ 
\hline  
  Fe$_{80}$ B$_{20}$                      &   40  & 320, 000  & $\lambda_{s}  \sim 30. \times 10^{-6}$ \\
  Fe$_{81}$ Si$_{3.5}$ B$_{13.5}$ C$_{2}$ &   43.7 & 260, 000    &    \\
\hline
  Fe$_{40}$ Ni$_{40}$ P$_{14}$ B$_{6}$    &   7.5 & 400, 000  &   $\lambda_{s}  \sim 10. \times 10^{-6}$\\
  Fe$_{40}$ Ni$_{38}$ Mo$_{4}$ B$_{18}$      &   12.5-50 & 200, 000 &  \\
  Fe$_{39}$ Ni$_{39}$ Mo$_{4}$ Si$_{6}$ B$_{12}$ &   12.5-50 & 200, 000 & \\
\hline
  Co$_{58}$ Ni$_{10}$ Fe$_{5}$ (Si, B)$_{27}$ &   10-12.5 & 200, 000 & $\lambda_{s}  \sim 0.1 \times 10^{-6}$ \\
  Co$_{66}$ Fe$_{4}$ (Mo, Si, B)$_{30}$ &   2.5-5 & 300, 000 &  \\
\hline  
\end{tabular}
\end{center}
\caption{Examples of soft magnetic materials and their hierarchy according to $\lambda_{s}$. The 
main composition is  successively based on Fe,  NiFe and finally Co. A commercial compound, close to the last family,
Vitrovac 6025~\textregistered or Co$_{66}$~Fe$_{4}$~Mo$_{2}$~Si$_{16}$~B$_{12}$  has  \cite{murillo} $\lambda_{s}=-1.4 \hspace{1mm} 10^{-7}$}
\label{tab2}
\end{table}

\begin{table}[!h]
\begin{center}
\begin{tabular}{|c|c|c|c|c|}
\hline
 Material & max(${r(H_{dc})}$) \footnote{$r(H_{dc})=100 \times |Z(H_{dc})-Z(H_{sat})|/|Z(H_{sat})|$} & max($dr(H_{dc})/dH_{dc}$) (\% /Oe) & Frequency (MHz) \\
\hline
Amorphous microwire: &  &   &  \\
Co$_{68.15}$ Fe$_{4.35}$ Si$_{12.5}$ B$_{15}$  & 56  &   58.4   & 0.9 \\
\hline
Finemet wire \footnote{Fe-Cu-Nb-B-Si family}  & 125  &    &  4 \\
\hline
Amorphous wire:  &       &  &  \\
Co$_{68.15}$ Fe$_{4.35}$ Si$_{12.5}$ B$_{15}$   & 220  &  1760 & 0.09 \\
\hline
CoP multilayers  &  &   &  \\
electroplated on Cu wire  & 230  &   & 0.09 \\
\hline
Mumetal \footnote{77\% Ni, 16\% Fe,  5\% Cu,  2\% Cr},  stripe &  310 & 20.8 & 0.6  \\
\hline
Textured      &        &       & \\
Fe-3\% Si  sheet &  360 &    & 0.1 \\
\hline
 Amorphous ribbon:                   &                &  &  \\
 Co$_{68.25}$ Fe$_{4.5}$ Si$_{12.25}$ B$_{15}$   & 400  &  &  1 \\
\hline
Ni$_{80}$ Fe$_{20}$ electroplated   &            &   &  \\
on non-magnetic CuBe microwire &     530 & 384 & 5 \\
\hline
Sandwich film: &    &   &  \\
CoSiB/SiO$_{2}$/Cu/SiO$_{2}$/CoSiB  &  700  & 304  & 20 \\
\hline
FeCoNi electroplated  &  &      &  \\
 on non-magnetic  CuBe microwire &    800 &   & 1.5 \\
\hline  
\end{tabular}
\end{center}
\caption{Materials for GMI sensors: wires and films adapted from Hauser et al. \cite{hauser}. 
The GMI ratio is given in the second column whereas the sensitivity with respect to the magnetic field 
is given by the largest value of the derivative of the ratio (third column).}
\label{tab3}
\end{table}

\begin{table}[!h]
\begin{tabular}{|l|c|c|}
\hline
     HL quantity & Head core&  Recording Media\\
\hline   
   $M_{s}$ &  Large &  Moderate\\
   $M_{r}$ &  Small &  Large\\
   $H_{c}$ &  Small &  Large\\
   $S= M_{r}/M_{s}$  &  Small $\sim 0$ &  $\sim 1$\\
  $S_{c}={(\frac{dM}{dH}) }_{ H=H_{c} }$ & Large &     \\
   $S^{*} = 1-\frac{ (M_{r}/H_{c}) }{S_{c}} $ &   &  $\sim 1$ \\
\hline     
\end{tabular}
\caption{ Requirements on read head and recording media from HL. The parameters S and S* are
the remanant and the loop squareness respectively. S* represents the steepness of the HL at the coercive
 field,  i.e. the slope of the HL at $H_{c}$ is $S_{c}={(\frac{dM}{dH}) }_{ H=H_{c} }=\frac{M_{r}}{H_{c}(1-{S}^{*})}$. 
In practice a fairly square HL has S,  S* $\sim 0.85$.}
\label{tab4}
\end{table}

\begin{table}[!h]
\begin{center}
\begin{tabular}{|c|c|c|}
\hline
 Magnetic induction \footnote{magnetic field is expressed in Oe and induction in Gauss.} occurrence &  Typical Range  &  Typical Range \\
\hline
 Type of Induction &   (in Gauss) &     (in Tesla) \\
\hline
\hline
Biological    &       &  \\
\hline
Body or brain of man \footnote{The magnetic induction of the Human Brain is about 10$^{-13}$ Tesla},  animals... & 10$^{-10}$ -  10$^{-5}$ Gauss  &  10$^{-14}$ - 10$^{-9}$ Tesla 
 \footnote{Human  heart induction is  around
10$^{-10}$ Tesla}  \\
\hline
\hline
Geological &     &  \\
\hline
Inside Earth  & 10$^{10}$  Gauss &  10$^{6}$  Tesla  \\
On the Earth surface  & 10$^{-4}$ - 1 Gauss \footnote{Transverse component of the Earth magnetic induction is
0.5 Gauss} &  \\
\hline
\hline
Superconducting Coil  & $>$ 25. 10$^{4}$ Gauss &  $>$ 25 Teslas \\
\hline  
\end{tabular}
\end{center}
\caption{Natural and artificial magnetic induction in various systems. Note that magnetic field unit is A/m and Oe  and 
Induction is Tesla  and Gauss. In vacuum or non-magnetic materials in CGS values in Oe and Gauss are same.}
\label{tab5}
\end{table}

\begin{table}[!h]
\begin{center}
\begin{tabular}{|c|c|c|}
\hline
 Sensor type  &  Magnetic induction typical range &  typical sensitivity \\
              &   (in Gauss)                  &            (in Gauss) \\
\hline
Hall              &   1 - 10$^{6}$  &    10. \\
Magnetoresistance  & 1 - 10$^{6}$  &  1.   \\
\hline
Magnetoimpedance    &   10$^{-6}$ - 1   &    10$^{-6}$ \\
Flux Gate &  10$^{-6}$ - 1 &    10$^{-6}$  \\
\hline
SQUID  & 10$^{-9}$ - 10$^{-6}$         &    10$^{-10}$  \\
\hline  
\end{tabular}
\end{center}
\caption{Magnetic induction sensor types,  range and typical sensitivity. Note that magnetic field unit is A/m and Oe  and 
induction is Tesla  and Gauss. In vacuum or non-magnetic materials,  CGS values in Oe and Gauss are same.}
\label{tab6}
\end{table}

%\end{article}

\end{document}